\RequirePackage{fixltx2e}
\documentclass[nofootinbib,reprint,showpacs,noeprint,
  prc,aps,10pt,superscriptaddress,altaffilletter,floatfix,longbibliography]{revtex4-1}
\pdfoutput=1
\usepackage{graphicx}
\usepackage{hyperref}
\usepackage{epstopdf}
\usepackage{amsmath}
\usepackage{amssymb}
\usepackage[capitalise]{cleveref}
\usepackage{ifthen}
\usepackage{booktabs}
\usepackage{dcolumn}
\newcolumntype{C}[1]{>{\centering\let\newline\\\arraybackslash\hspace{0pt}}m{#1}}

\graphicspath{{figs/}}

\begin{document}

\newcommand{\isot}[2]{$^{#2}$#1}
\newcommand{\isotbold}[2]{$^{\boldsymbol{#2}}$#1}
\newcommand{\xeiso}{\isot{Xe}{136}}
\newcommand{\thsrc}{\isot{Th}{228}}
\newcommand{\cosrc}{\isot{Co}{60}}
\newcommand{\rasrc}{\isot{Ra}{226}}
\newcommand{\cssrc}{\isot{Cs}{137}}
\newcommand{\betascale}  {$\beta$-scale}
\newcommand{\kevkgyr}  {keV$^{-1}$ kg$^{-1}$ yr$^{-1}$}
\newcommand{\nonubb}  {$0\nu \beta \beta$}
\newcommand{\nonubbbf}  {$\boldsymbol{0\nu \beta \beta}$}
\newcommand{\twonubb} {$2\nu \beta \beta$}
\newcommand{\vadc} {ADC$_\text{V}$}
\newcommand{\uadc} {ADC$_\text{U}$}
\newcommand{\mus} {\textmu{}s}
\newcommand{\chisq} {$\chi^2$}
\newcommand{\mum} {\textmu{}m}
\newcommand{\checkit}[1]{{\color{red}#1}}
\newcommand{\RunTwoA}{Run 2a}
\newcommand{\RunTwo}{Run 2}
\newcommand{\RunTwoBC}{Runs 2b and 2c}
\newcommand{\SP}[1]{\textsuperscript{#1}}
\newcommand{\SB}[1]{\textsubscript{#1}}
\newcommand{\SPSB}[2]{\rlap{\textsuperscript{#1}}\SB{#2}}
\newcommand{\pmasy}[3]{#1\SPSB{$+$#2}{$-$#3}}
\newcommand{\matel}{$M^{2\nu}$}
\newcommand{\psfac}{$G^{2\nu}$}
\newcommand{\tbeta}{T$_{1/2}^{0\nu\beta\beta}$}
\newcommand{\exolimit}[1][true]{\pmasy{2.6}{1.8}{2.1}$ \cdot 10^{25}$}
\newcommand{\exomeasurement}{\tbeta{}= \exolimit{}~yr}
\newcommand{\U}{\text{U}}
\newcommand{\V}{\text{V}}
\newcommand{\X}{\text{X}}
\newcommand{\Y}{\text{Y}}
\newcommand{\Z}{\text{Z}}
\newcommand{\bqcm}{${\rm Bq~m}^{-3}$}
\newcommand{\nonunorm}{N_{{\rm Err, } 0\nu\beta\beta}}
\newcommand{\nonunum}{n_{0\nu\beta\beta}}
\newcommand{\cussim}[1]{$\sim$#1}
\newcommand{\halflife}[1]{$#1\cdot10^{25}$~yr}
\newcommand{\numspec}[3]{$N_{^{#2}\mathrm{#1}}=#3$}

\title{Search for Majorana neutrinos with the first two years of EXO-200 data}

\newcommand{\IHEP}{\affiliation{Institute of High Energy Physics, Beijing, China}}
\newcommand{\DukeTUNL}{\affiliation{Department of Physics, Duke University, and Triangle Universities Nuclear Laboratory (TUNL), Durham, North Carolina, USA}}
\newcommand{\TRIUMF}{\affiliation{TRIUMF, Vancouver, BC, Canada}}
\newcommand{\Seoul}{\affiliation{Department of Physics, University of Seoul, Seoul, Korea}}
\newcommand{\Stanford}{\affiliation{Physics Department, Stanford University, Stanford CA, USA}}
\newcommand{\WIPP}{\affiliation{Waste Isolation Pilot Plant, Carlsbad NM, USA}}
\newcommand{\Laurentian}{\affiliation{Department of Physics, Laurentian University, Sudbury ON, Canada}}
\newcommand{\ITEP}{\affiliation{Institute for Theoretical and Experimental Physics, Moscow, Russia}}
\newcommand{\Illinois}{\affiliation{Physics Department, University of Illinois, Urbana-Champaign IL, USA}}
\newcommand{\Caltech}{\affiliation{Kellogg Lab, Caltech, Pasadena, CA, USA}}
\newcommand{\UMass}{\affiliation{Physics Department, University of Massachusetts, Amherst MA, USA}}
\newcommand{\TUM}{\affiliation{Technische Universit\"at M\"unchen, Physikdepartment and Excellence Cluster Universe, Garching, Germany}}
\newcommand{\Maryland}{\affiliation{Physics Department, University of Maryland, College Park MD, USA}}
\newcommand{\LHEP}{\affiliation{LHEP, Albert Einstein Center, University of Bern, Bern, Switzerland}}
\newcommand{\CSU}{\affiliation{Physics Department, Colorado State University, Fort Collins CO, USA}}
\newcommand{\Indiana}{\affiliation{Physics Department and CEEM, Indiana University, Bloomington IN, USA}}
\newcommand{\Alabama}{\affiliation{Department of Physics and Astronomy, University of Alabama, Tuscaloosa AL, USA}}
\newcommand{\Drexel}{\affiliation{Department of Physics, Drexel University, Philadelphia PA, USA}}
\newcommand{\SLAC}{\affiliation{SLAC National Accelerator Laboratory, Stanford CA, USA}}
\newcommand{\Carleton}{\affiliation{Physics Department, Carleton University, Ottawa ON, Canada}}
\author{J.B.~Albert}\Indiana
\author{D.J.~Auty}\Alabama
\author{P.S.~Barbeau}\DukeTUNL
\author{E.~Beauchamp}\Laurentian
\author{D.~Beck}\Illinois
\author{V.~Belov}\ITEP
\author{C.~Benitez-Medina}\CSU
\author{J.~Bonatt}\UMass\Stanford
\author{M.~Breidenbach}\SLAC
\author{T.~Brunner}\Stanford
\author{A.~Burenkov}\ITEP
\author{G.F.~Cao}\IHEP
\author{C.~Chambers}\CSU
\author{J.~Chaves}\Stanford
\author{B.~Cleveland}\altaffiliation{Also SNOLAB, Sudbury ON, Canada}\Laurentian
\author{M.~Coon}\Illinois
\author{A.~Craycraft}\CSU
\author{T.~Daniels}\UMass
\author{M.~Danilov}\ITEP
\author{S.J.~Daugherty}\Indiana
\author{C.G.~Davis}\Maryland
\author{J.~Davis}\Stanford
\author{R.~DeVoe}\Stanford
\author{S.~Delaquis}\LHEP
\author{T.~Didberidze}\Alabama
\author{A.~Dolgolenko}\ITEP
\author{M.J.~Dolinski}\Drexel
\author{M.~Dunford}\Carleton
\author{W.~Fairbank Jr.}\CSU
\author{J.~Farine}\Laurentian
\author{W.~Feldmeier}\TUM
\author{P.~Fierlinger}\TUM
\author{D.~Fudenberg}\Stanford
\author{G.~Giroux}\altaffiliation{Now at Queen's University, Kingston, ON, Canada}\LHEP
\author{R.~Gornea}\LHEP
\author{K.~Graham}\Carleton
\author{G.~Gratta}\Stanford
\author{C.~Hall}\Maryland
\author{S.~Herrin}\SLAC
\author{M.~Hughes}\Alabama
\author{M.J.~Jewell}\Drexel
\author{X.S.~Jiang}\IHEP
\author{A.~Johnson}\SLAC
\author{T.N.~Johnson}\Indiana
\author{S.~Johnston}\UMass
\author{A.~Karelin}\ITEP
\author{L.J.~Kaufman}\Indiana
\author{R.~Killick}\Carleton
\author{T.~Koffas}\Carleton
\author{S.~Kravitz}\Stanford
\author{A.~Kuchenkov}\ITEP
\author{K.S.~Kumar}\UMass
\author{D.S.~Leonard}\Seoul
\author{F.~Leonard}\Carleton
\author{C.~Licciardi}\Carleton
\author{Y.H.~Lin}\Drexel
\author{R.~MacLellan}\SLAC
\author{M.G.~Marino}\email[Corresponding author: ]{michael.marino@mytum.de}\TUM
\author{B.~Mong}\Laurentian
\author{D.~Moore}\Stanford
\author{R.~Nelson}\WIPP
\author{A.~Odian}\SLAC
\author{I.~Ostrovskiy}\Stanford
\author{C.~Ouellet}\Carleton
\author{A.~Piepke}\Alabama
\author{A.~Pocar}\UMass
\author{C.Y.~Prescott}\SLAC
\author{A.~Rivas}\Stanford
\author{P.C.~Rowson}\SLAC
\author{M.P.~Rozo}\Carleton
\author{J.J.~Russell}\SLAC
\author{A.~Schubert}\Stanford
\author{D.~Sinclair}\Carleton\TRIUMF
\author{S.~Slutsky}\altaffiliation{Now at California Institute of Technology, Pasadena, CA, USA}\Maryland
\author{E.~Smith}\Drexel
\author{V.~Stekhanov}\ITEP
\author{M.~Tarka}\Illinois
\author{T.~Tolba}\LHEP
\author{D.~Tosi}\altaffiliation{Now at University of Wisconsin, Madison, WI, USA}\Stanford
\author{K.~Twelker}\Stanford
\author{P.~Vogel}\Caltech
\author{J.-L.~Vuilleumier}\LHEP
\author{A.~Waite}\SLAC
\author{J.~Walton}\Illinois
\author{T.~Walton}\CSU
\author{M.~Weber}\Stanford
\author{L.J.~Wen}\IHEP
\author{U.~Wichoski}\Laurentian
\author{J.D.~Wright}\UMass
\author{L.~Yang}\Illinois
\author{Y.-R.~Yen}\Maryland\Drexel
\author{O.Ya.~Zeldovich}\ITEP
\author{Y.B.~Zhao}\IHEP

\collaboration{EXO-200 Collaboration}

\date{\today}

\begin{abstract}

Many extensions of the Standard Model of particle physics suggest that
neutrinos should be Majorana-type fermions, but this assumption is difficult to
confirm.  Observation of neutrinoless double-beta decay (\nonubb{}), a
spontaneous transition that may occur in several candidate nuclei, would verify
the Majorana nature of the neutrino and constrain the absolute scale of the
neutrino mass spectrum.  Recent searches carried out with \isot{Ge}{76} (GERDA
experiment) and \isot{Xe}{136} (KamLAND-Zen and EXO-200 experiments) have
established the lifetime of this decay to be longer than $10^{25}$~yr,
corresponding to a limit on the neutrino mass of 0.2\textendash{}0.4~eV.  Here
we report new results from EXO-200 based on 100~kg$\cdot$yr of \isot{Xe}{136}
exposure, representing an almost fourfold increase from our earlier published
datasets.  We have improved the detector resolution at the \isot{Xe}{136}
double-beta-decay Q-value to $\sigma/E = 1.53\%$ and revised the data analysis.
The obtained half-life sensitivity is \halflife{1.9}, an improvement by a
factor of 2.7 compared to previous EXO-200 results.  We find no statistically
significant evidence for \nonubb{} decay and set a half-life limit of
\halflife{1.1} at 90\% CL.  The high sensitivity holds promise for further
running of the EXO-200 detector and future \nonubb{} decay searches with nEXO. 

\end{abstract}


\maketitle

Majorana fermions, a class of neutral spin-1/2 particles described by 2-component spinors,
have been an element of quantum field theory since its
inception~\cite{Majorana:1937vz,Racah:1937qq}.  Electrons and other
spin-1/2 elementary particles with distinct antiparticles, however, are described by
4-component Dirac spinors.  Majorana quasiparticles may have
been observed in condensed matter systems~\cite{Mourik25052012} where
neutrality is achieved through the collective action of electrons and holes.
Among the known {\it elementary} particles, only neutrinos are 
Majorana fermion candidates, owing to their intrinsic neutrality.  Confirmation of
this property would imply the non-conservation of lepton number, an
additive quantum number that, unlike charge or color, is not related to any
known gauge symmetry. As yet, lepton number has been empirically found to be conserved.
Neutrinos are also remarkable for their small, yet finite,
masses~\cite{Camilleri:2008zz} that are generally difficult to explain, but
arise naturally in many extensions~\cite{Mohapatra:1979ia,GellMann:1980vs} of
the Standard Model of particle physics (SM).  A generic consequence of many
such extensions is that neutrinos should be of the Majorana variety. 

The most sensitive probe for Majorana neutrinos is a nuclear process known as
neutrinoless double-beta decay (\nonubb{}), whereby a nucleus decays by
emitting two electrons and nothing else, while changing its charge by two
units~\cite{Schechter:1981bd}.  A related double-beta decay process, known as
two-neutrino double-beta decay (\twonubb{}), is allowed by the SM and has been
observed in many nuclei, \isot{Xe}{136} among them~\cite{Barabash:2010ie,Albert2013}.  It provides,
however, no direct information on the Majorana/Dirac question.  The exotic
\nonubb{} can be distinguished from the \twonubb{} by measuring the sum energy
of the two electrons that is peaked at the Q-value for the former and is a
continuum for the latter.
We refer to this region around the Q-value as the \nonubb{} region-of-interest (ROI).
The half-life of the \nonubb{} is related to the effective Majorana
neutrino mass ($\langle m_{\beta\beta} \rangle$) by a phase space factor and a
nuclear matrix element.   Hence the observation of the \nonubb{} decay would
discover elementary Majorana particles, demonstrate lepton number violation and
measure the neutrino mass scale $\langle m_{\beta\beta} \rangle$, at least to
within the theoretical uncertainty of the nuclear matrix
elements~\cite{Vogel:2012ja}.

Recent sensitive searches for \nonubb{} have been carried out in \isot{Ge}{76}
(GERDA~\cite{Agostini:2013mzu}) and \isot{Xe}{136} (KamLAND-Zen~\cite{Gando:2012zm} and
EXO-200~\cite{Auger:2012ar}). These experiments
have set limits on the Majorana neutrino mass of \cussim{0.2\textendash{}0.4~eV}, and
have 
cast doubt on an earlier claim of observation~\cite{KlapdorKleingrothaus:2006ff}. In this letter we
report on new \nonubb{} search results from the EXO-200 experiment 
based upon about two years of data.

\section{EXO-200 detector}
EXO-200 has been described in detail elsewhere~\cite{Auger:2012gs}.  Briefly,
the detector is a cylindrical liquid xenon (LXe) time projection chamber (TPC),
roughly 40~cm in diameter and 44~cm in length. Two drift regions are separated
in the center by a cathode. The LXe is enriched to 80.6\% in \isot{Xe}{136}, the
\nonubb{} candidate  ($Q=2457.83\pm 0.37$~keV~\cite{Redshaw:2007}).  The TPC
provides X-Y-Z coordinate and energy measurements of ionization deposits in the
LXe by simultaneously collecting the scintillation light and the charge.
Charge deposits spatially separated by about 1~cm or more are individually
observed and the position accuracy for isolated deposits is a few mm.
Avalanche Photodiodes (APDs) measure the scintillation light.  Small
radioactive sources can be positioned at standard positions near the TPC to
calibrate the detector and monitor its stability.

The TPC is shielded from environmental radioactivity on all sides by
\cussim{50~cm} of HFE-7000 cryofluid~\cite{3m} (HFE) maintained at \cussim{167~K} inside a
vacuum-insulated copper cryostat.  Further shielding is provided by at least 25~cm of
lead in all directions. The entire assembly is housed in a
cleanroom located underground at a depth of $1585^{+11}_{-6}$~meters water
equivalent~\cite{Esch:2004zj} at the Waste Isolation Pilot Plant near Carlsbad,
NM, USA. Four of the six sides of the cleanroom are instrumented with plastic
scintillator panels recording the passage of cosmic ray muons.  An extensive
materials screening campaign~\cite{Leonard:2007uv} was employed to minimize the
radioactive background produced by the detector components.

\section{Data analysis and methodology}
The data analysis methods in this work follow closely those presented in detail
in~\cite{Albert2013}.  Events in the detector are classified as single-site
(SS) or multi-site (MS) according to the number of detected charge deposits.
\nonubb{} events are predominantly SS whereas $\gamma$ backgrounds are mostly MS. 
For each
event, the energy is determined as a linear combination of charge and scintillation, while a
``standoff distance'' (SD) is defined as the distance between a charge deposit
and the closest material that is not LXe, other than the cathode.   To search for \nonubb{},
a binned maximum-likelihood (ML) fit is performed simultaneously over the SS
and MS events using probability density functions (PDFs) in energy and SD,
generated using a Geant4-based~\cite{GEANT42006} Monte Carlo simulation (MC). The energy range
980\textendash{}9800~keV is used.  The `low-background data set' (physics data)
is obtained after applying event selection cuts.   With respect to
Ref.~\cite{Albert2013} the current analysis additionally includes: (1)~improved
signal processing for the scintillation waveforms resulting in lower noise;
(2)~\rasrc{} source calibration data; (3)~an expanded fiducial volume; (4)~the
estimation of systematic errors related to the \nonubb{} ROI; and (5)~updated
background and systematic studies relevant to the \nonubb{} search. 

The data set presented here (\RunTwo{}) combines \RunTwoA{} (already used
for~\cite{Auger:2012ar,Albert2013}, September 22, 2011 \textendash{} April 15,
2012) and \RunTwoBC{} (April 16, 2012 \textendash{} September 1, 2013).  After
removing periods of poor data quality and calibration runs, the total amount of
low-background data for this analysis is $477.60\pm 0.01$~days, a
3.8-fold increase from previous EXO-200 publications.  The primary
tool used for understanding and correcting the detector energy measurement is
the 2615~keV $\gamma$ line of \isot{Tl}{208} from a \thsrc{} source deployed at
least twice weekly during the time spanned by this data set.  Seven multi-day
calibration campaigns involving the use of multiple sources (\thsrc{},
\cosrc{},
\rasrc{} and \cssrc{}) were performed at roughly 3-month intervals throughout
the data set.  The lifetime of ionization electrons in the LXe is better than
2~ms for the entire data set, more than sufficient to collect charge across the
full volume of the detector.  We determine the optimal linear combination of
scintillation and ionization signals once per week by minimizing the width of
the 2615~keV line. To prevent making analysis decisions that could bias the
results in the ROI, the low-background data were partially ``masked" to hide \cussim{2/3}
of the live-time for SS events between 2325 and 2550~keV.  Live-time already
analyzed in previous publications (e.g.~\RunTwoA{}) was not masked.  

\begin{figure}
	\includegraphics[width=0.48\textwidth]{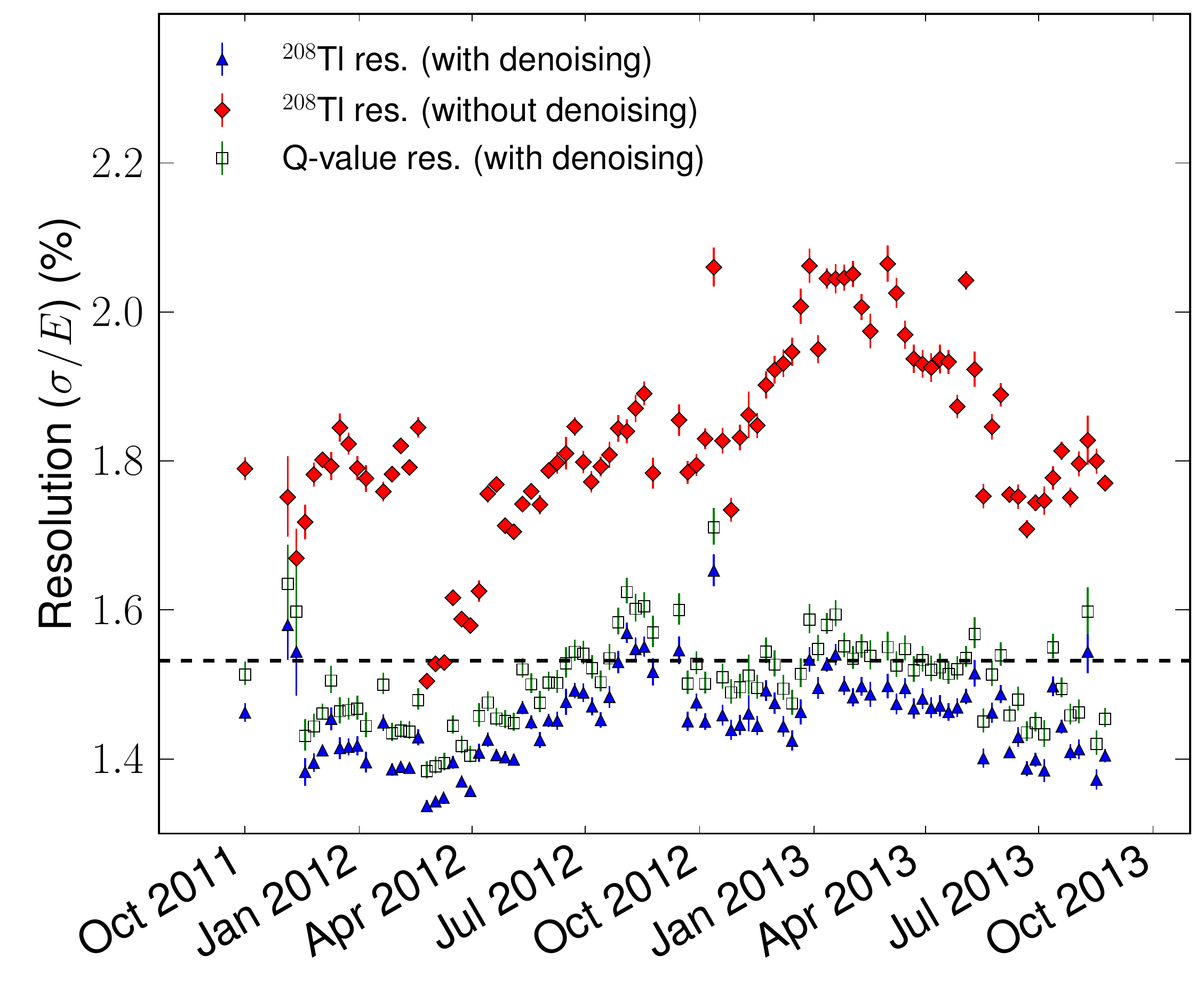}
	\caption{\textbf{Effect of denoising on the energy resolution
($\boldsymbol{\sigma/E}$)}.  Resolution for SS events versus time with and without
application of the denoising algorithm.  Shown are resolutions at the 2615~keV
\isot{Tl}{208} full-absorption peak (with and without) and propagated to the \nonubb{}
Q-value (with). The time variation is caused by changes in the noise of the APD
front-end electronics.  The horizontal dashed line shows the effective Q-value
SS energy resolution used for the data set (1.53\%).  MS resolution (not shown)
exhibits similar behavior.  The error bars represent $\pm1$ standard deviation intervals. }
    \label{fig:Denoising}
\end{figure}

The energy resolution of the detector is dominated by electronic noise in the
scintillation readout and exhibits variations over time due to changes in
this noise.  We apply a denoising algorithm to the scintillation signals during
post processing, improving the detector resolution and reducing its time
dependence.  This algorithm attempts to find the optimal combination of APD
waveforms to determine the amount of scintillation light for each event, taking into account the
measured electronic noise of each APD channel as well as the position of each
charge deposition in the detector.  \Cref{fig:Denoising} shows the resolution
with and without denoising.

We define an effective, time-independent energy resolution
function~\cite{Albert2013} $\sigma^{2}(E)=\sigma^2_{\rm elec}+b^{2}E+c^{2}E^{2}$.  
Here, $\sigma_{\rm elec}$, $b$ and $c$ are 20.8~keV, 0.628~keV$^{1/2}$ and
$1.10\cdot10^{-3}$ (25.8~keV, 0.602~keV$^{1/2}$ and $4.04\cdot10^{-3}$) for SS
(MS), determined by a ML fit to calibration data taken during \RunTwo{}.  This
function is folded with the energy distributions derived from the simulation to create the PDFs used in final fits.
The effective resolution ($\sigma$/E) for SS (MS) at the \nonubb{} Q-value is
$1.53\pm0.06\%$ ($1.65\pm0.05\%$).

The fiducial volume (FV) is larger than in~\cite{Albert2013} 
to maximize the sensitive mass while maintaining systematic uncertainties at
an acceptable level.  Events in the FV are required to have 182~mm $>|\Z|>$ 10~mm (where $\Z=0$ 
is the cathode plane) and are contained in a hexagon with 162~mm apothem.
This represents a \isot{Xe}{136} mass of 76.5~kg, corresponding to
$3.39\times10^{26}$ atoms of \isot{Xe}{136} and, with the quoted live-time,
results in an exposure of 100~kg$\cdot$yr (736~mol$\cdot$yr).

\section{Investigation and determination of systematic errors}
The main systematic uncertainties relevant to the search for \nonubb{} 
are related to (1)~signal efficiency, (2)~location of the \nonubb{} ROI within
the spectrum, and (3)~estimation of the background in the ROI.  

\begin{figure}
    \includegraphics[width=0.48\textwidth]{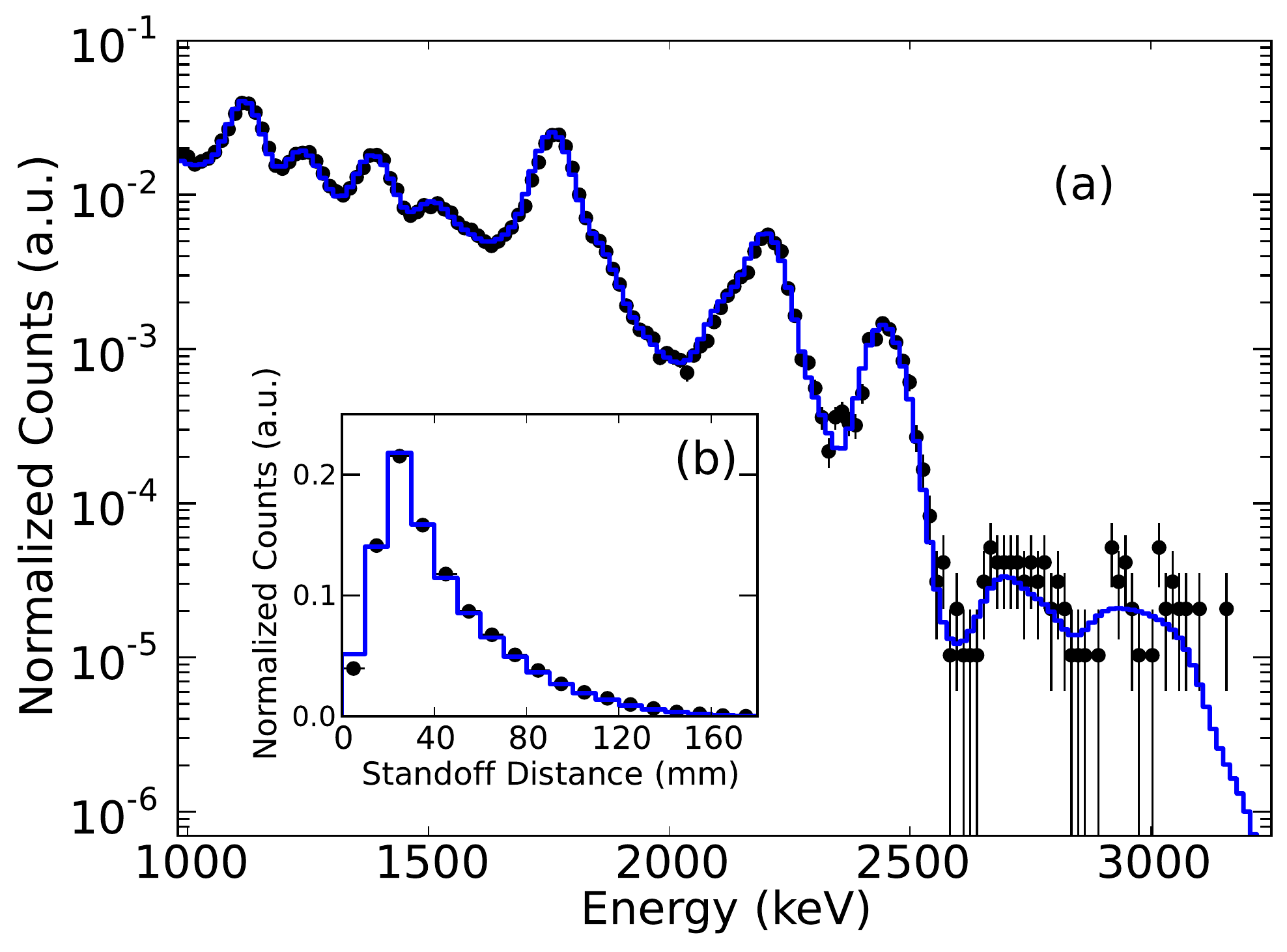}
	\caption{\textbf{Comparison of energy and SD distributions of a \isotbold{Ra}{226} calibration source for simulation
and data.} Energy (a) and SD (b) are shown for data (black points) and
simulation (blue line).  The calibration source is at a position near the cathode
outside the TPC.  The error bars represent $\pm1$ standard deviation intervals. }

    \label{fig:RaSrcCompare}
    
\end{figure}

To verify the simulation's ability to model efficiencies and the
background, we compare measurement and simulation of calibration
sources deployed at various positions around the TPC, investigating in
particular: (a)~the energy and SD distributions, (b)~the integrated
rate of selected events, and (c)~the SS/MS event ratio versus energy.   
A representative set of results for (a) is shown in \cref{fig:RaSrcCompare},
where simulation-data agreement for the \rasrc{} source are presented.  
\rasrc{} is a particularly valuable source because of several $\gamma$
lines that map a broad energy region including the \nonubb{} ROI. The energy
spectrum shows good agreement across the energy range of the analysis.
Comparable results were also obtained with the \cosrc{} and \thsrc{} sources.
The SD agreement is within statistical errors except in the first 10~mm bin,
where the simulation produces more events in the FV than seen in data.

Discrepancies in the shapes of energy and SD distributions between data and
simulation affect the estimation of the background in the \nonubb{} ROI.   
To quantify this effect, we calculate skewing functions based upon the small
discrepancies observed in source calibration studies.  We distort the
background PDFs with the skewing functions and use these to produce a set
of toy MC data sets.  The toy MC data sets are then fit to un-skewed PDFs.  The
change in the \nonubb{} ROI background is 9.2\%, which we take as systematic
error.

In the rate comparison studies (b), we combine the total number of selected
events in data and simulation as $({\rm Data} - {\rm MC})/{\rm Data}$ for
several source positions.  The error-weighted average of the results
is calculated using the FV in this
analysis as well as the FV in~\cite{Albert2013}.  The difference between these
values is 1.7\%, which we combine with the underlying FV uncertainty (also 1.7\%,
\cite{Albert2013}) conservatively assuming full correlation to produce a total
error on the detector efficiency of 3.4\%. 

\begin{figure}
    \includegraphics[width=0.48\textwidth]{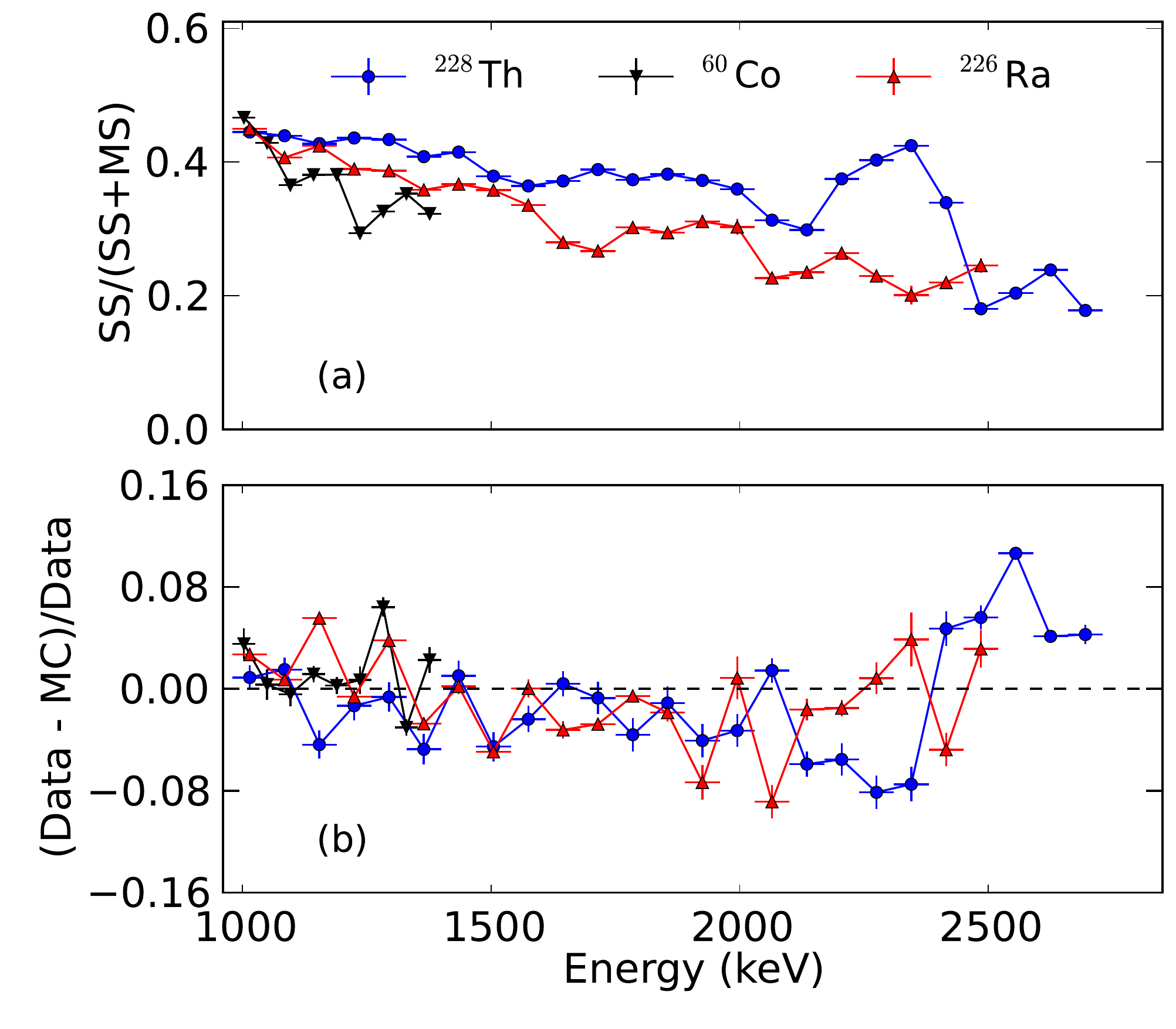}
	\caption{\textbf{Event multiplicity in data and simulation.} (a) SS/(SS+MS)
ratio in data for \rasrc{}, \cosrc{} and \thsrc{} calibration sources.  (b)
Comparison of SS/(SS+MS) ratio between data and simulation for the three
sources. Despite having different underlying energy spectra, all sources
exhibit similar behavior across the shown energy range when comparing data and
simulation (b). The error bars represent $\pm1$ standard deviation intervals. }
    \label{fig:AllSrcSSMS}
\end{figure}

The ratio of the number of SS events to the total number of events (SS/(SS+MS))
is compared between data and simulation for three sources in \cref{fig:AllSrcSSMS}. 
The general behavior is largely independent of the underlying spectral shape.
We choose to assign a single systematic uncertainty to the SS/(SS+MS) ratio of
9.6\%, calculated from the weighted average of the maximum deviations observed
for the \thsrc{}, \cosrc{} and \rasrc{} (data from the latter available after
June 2013) sources at several different source locations in each calibration
campaign.  

Event selection requires an event to be fully reconstructed in all
3~coordinates (X, Y and Z).  We compare the relative efficiency of this
requirement for \twonubb{} from MC to the measured relative efficiency derived
from the background-subtracted low-background energy spectrum.  Here, we define
the relative efficiency as the ratio of the number of events passing the entire
set of selection requirements to the number passing the set \emph{not
including} the full-reconstruction requirement.  The relative efficiency from
simulation changes modestly across the \twonubb{} energy range ($>99\%$ to
$90\%$ from 980~keV to 2450~keV) and similar behavior is seen in data.  The
average deviation between simulation and data over the \twonubb{} spectrum
(7.8\%) is taken as a systematic error on the efficiency. 

The uncertainty on the location of the ROI in the spectrum is dominated by a
possible energy-scale difference between $\beta$-like events in the LXe
(e.g.~\nonubb{}) and $\gamma$-like events (including most backgrounds and the
sources used for the primary energy calibration).  We define the `\betascale{}'
as $\mathrm{E}_{\beta} = B\cdot\mathrm{E}_{\gamma}$, where $\mathrm{E}_{\beta}$
($\mathrm{E}_{\gamma}$) is the energy for depositions from $\beta$s ($\gamma$s)
and $B$ is a measured constant.  We determine the \betascale{} by fitting to
the \twonubb{}-decay-dominated low-background data and find $B =
0.999\pm0.002$.  

Several cross checks were performed to search for energy dependence in the
\betascale{}.  The above fits were performed using
different energy thresholds and with different background PDFs produced
using the skewing functions discussed earlier.  We also fit
the low-background data assuming a linear energy dependence (e.g.~$p_0 + p_1
\mathrm{E}_{\gamma}$) for $B$.  In all cases the results are consistent with the
original fit, providing no evidence for energy dependence of the \betascale{}.
The estimate of the \betascale{} is also robust against a different
choice of \twonubb{} spectral shape~\cite{Kotila:2012zza}. 

To investigate the dependence of the ROI background estimate on the
completeness of the fit model, we derive PDFs from different source locations and
introduce them separately
into the default background model used in the fit.  The relative change of the estimated ROI background is then
determined.  The three background PDFs considered in this study are
\isot{U}{238} in the HFE and inner cryostat, and \isot{Co}{60} in the copper
source guide tube.  These were chosen because the initial source location
affects relative amplitudes and spectral features in the ROI, i.e.~the
\isot{Bi}{214} $\gamma$ (2448~keV) and \isot{Co}{60} sum peak.  This study
indicates a total possible deviation of 5.7\% for the expected background
counts in the ROI.  

The residual time dependence of the energy resolution (\cref{fig:Denoising})
can introduce additional counts in the ROI from the 2615~keV
\isot{Tl}{208} peak.  This was estimated to affect the ROI background counts by
$\pm1.5\%$. 

A summary of the \nonubb{} signal efficiency and associated uncertainty is
presented in \cref{tab:syserrors}.  \Cref{tab:bkgsyserrors} summarizes 
the uncertainties on the estimation of background in the ROI.  These errors are
explicitly included as input to the final fit to the low-background data.  Items not
listed in the tables, such as the \betascale{} and the SS/MS ratios, still
contribute to the total systematic error on the \nonubb{} signal as they are
propagated to the final result by the ML fit to the low-background data. 

\begin{table}
\begin{ruledtabular}
    \renewcommand{\arraystretch}{1.3}
    \begin{tabular}{lrcD{.}{.}{-1}}
        \toprule
        \multicolumn{2}{l}{Source} & \multicolumn{1}{p{0.13\textwidth}}{\centering Signal eff. (\%)} & 
                                     \multicolumn{1}{p{0.10\textwidth}}{\centering Error (\%)} \\
        \colrule
        Event selections \\
        \multicolumn{2}{p{0.2\textwidth}}{\raggedleft Summary from~\cite{Albert2013}}    & 93.1  & 0.9 \\
        \multicolumn{2}{p{0.2\textwidth}}{\raggedleft Partial reconstruction }           & 90.9  & 7.8 \\
        \multicolumn{2}{p{0.2\textwidth}}{\raggedleft Fiducial Volume/Rate agreement }   & -     & 3.4 \\
        \hline
        \multicolumn{2}{l}{Total}                                        & 84.6  & 8.6 \\
    \end{tabular}
\end{ruledtabular}
   \caption{\textbf{\nonubbbf{} signal efficiency and associated systematic errors.}
`Partial reconstruction' refers to the
requirement that all events be fully reconstructed in X,Y and Z.  The summary
for event selection from~\cite{Albert2013} includes all efficiencies and
related errors except fiducial volume and partial reconstruction, which have
been recalculated in this work for \nonubb{}.  } 

\label{tab:syserrors} 

\end{table}

\begin{table}
\begin{ruledtabular}
    \renewcommand{\arraystretch}{1.3}
    \begin{tabular}{lD{.}{.}{-1}}
        \toprule
        Source & \multicolumn{1}{p{0.10\textwidth}}{\centering Error (\%)} \\
        \colrule
        Background shape distortion    & 9.2 \\
        Background model               & 5.7 \\
        Energy resolution variation    & 1.5 \\
        \hline
        Total                    & 10.9 \\
    \end{tabular}
\end{ruledtabular}
   \caption{\textbf{Systematic errors on the determination of background in the
ROI.}  These arise from incorrect modeling of the background shape (Background
shape distortion), incorrect or incomplete background model (Background model),
and the residual variation of the energy resolution over time (Energy
resolution variation, see e.g.~\cref{fig:Denoising}). 
} 
\label{tab:bkgsyserrors} 

\end{table}

\begin{figure*}
    \centering
    \includegraphics[width=0.98\textwidth]{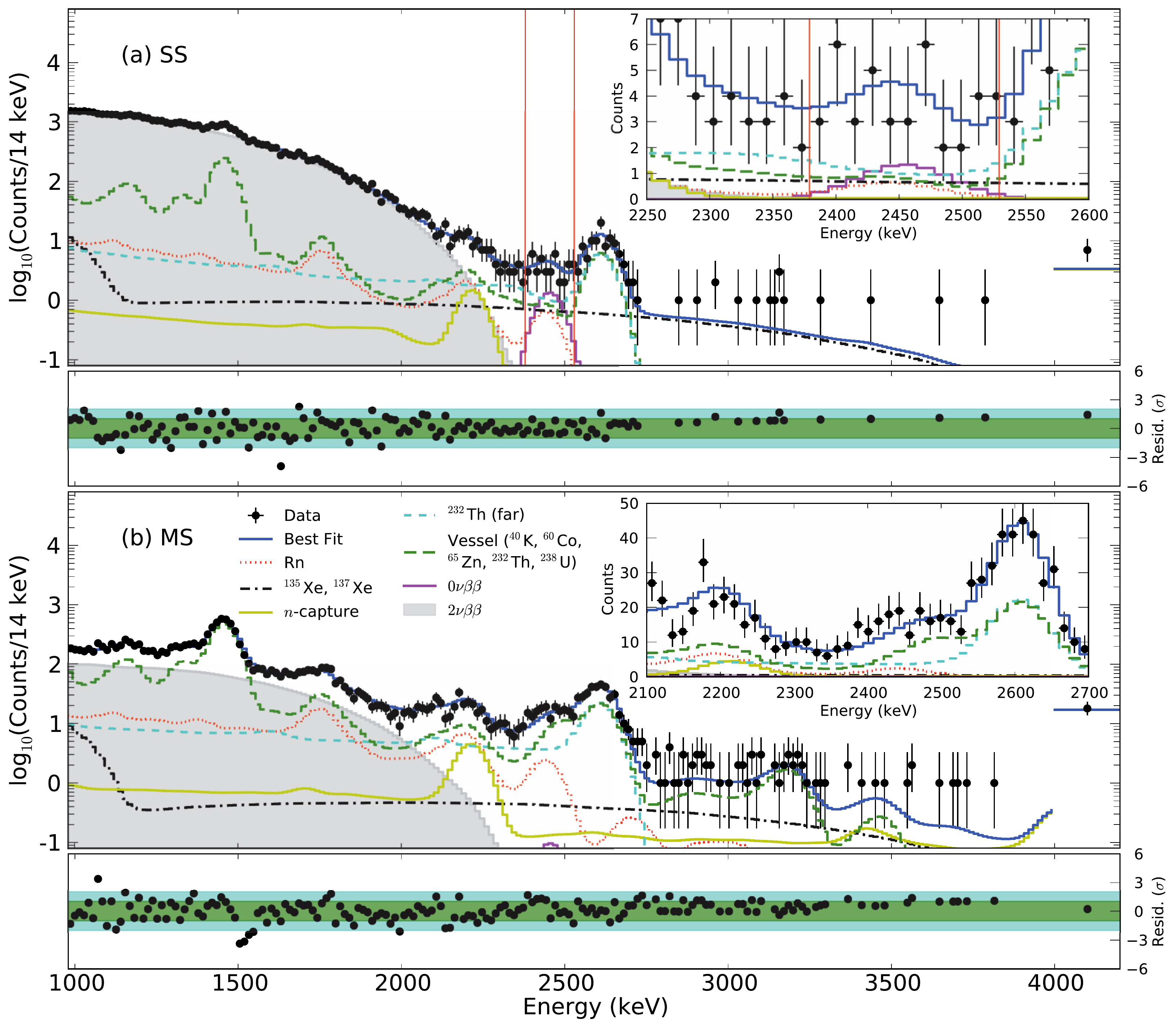}
	\caption{\textbf{Fit results projected in energy.} SS (a) and
MS (b) events are shown with a zoom-in (inset) around the ROI region:
2250\textendash{}2600~keV (2100\textendash{}2700~keV) for SS (MS).  The bin
size is 14~keV.  Data points are shown in black and residuals between data and
best fit normalized to the Poisson error are presented, ignoring bins with
0~events.  The 7 (18) events between 4000 and 9800~keV in the SS (MS) spectrum
have been collected into an overflow bin for presentation here.  The vertical
(red) lines in the SS spectra indicate the $\pm2\sigma$~ROI.  The result of the
simultaneous fit to the SD is not shown here. Several background model
components, including Rn, \isot{Xe}{135} and \isot{Xe}{137}, $n$-capture,
\isot{Th}{232} (far); Vessel; \nonubb{}; and \twonubb{} (described further in
the text), are indicated in main panel (b) to show their relative contributions
to the spectra. The error bars on the data points represent $\pm1$ standard
deviation intervals. } 
\label{fig:BestFit}
\end{figure*}

Neutrons arising from cosmic-ray muons or radioactive decays in the salt
surrounding the laboratory may contribute background to the \nonubb{} ROI via
neutron capture or spallation processes.  The contribution in the ROI is
expected to arise primarily from neutron-capture $\gamma$s in the
LXe and surrounding materials (e.g.~capture on \isot{Cu}{63} and \isot{Cu}{65} in
the copper components, and on \isot{Xe}{136} in the
LXe).  A simulation using a simplified experimental geometry and employing the
FLUKA~\cite{FLUKA1,FLUKA2} and SOURCES~\cite{SOURCES} software packages is
used to generate neutrons, track and thermalize them.  The resulting neutron
capture rates are used as input to the Geant4-based~\cite{GEANT42006} EXO-200
simulation package~\cite{Albert2013}, with the respective n-capture
$\gamma$-spectra produced based upon ENSDF information~\cite{ENSDF} for the
given nuclides.  The produced PDFs are used in fits to the low-background
data.  Good shape agreement is found between these PDFs and data coincident
with muon-veto-panel events.

\section{Results}
The fit to the low-background data minimizes the negative log-likelihood
function constructed using a signal and background model composed of PDFs
from simulation.  A profile-likelihood (PL) scan is performed to search for a
\nonubb{} signal. 

The PDFs chosen for the low-background fit model are those used
in~\cite{Albert2013} plus a ``far-source" \isot{Th}{232} PDF, a \isot{Xe}{137} PDF and
neutron-capture-related PDFs, including \isot{Xe}{136} neutron capture in the
LXe, \isot{H}{1} neutron-capture in the
HFE, and \isot{Cu}{63},\isot{Cu}{65} neutron capture in Cu components (LXe
vessel, inner and outer cryostats).  The far-source \isot{Th}{232} PDF allows for
background contributions from Th in materials far from the TPC, for example in
the HFE and in the copper cyrostat.  (Remote \isot{U}{238} is included in the
fit model via \isot{Rn}{222}, simulated in the air between the cryostat and Pb
shield.)  We combine the neutron-capture-related PDFs to form one PDF, allowing
the relative rates of the component PDFs to float within 20\% of their
simulation-estimated values.  The total rate of this summed PDF is allowed to
float unconstrained. 

We constrain the single-site fractions (SS/(SS+MS)) of all components to be
within 9.6\% of their value calculated from simulation.  
An additional 90\% correlation between single-site fractions of $\gamma$
components is introduced into the likelihood function, owing to the consistent
behavior observed in these parameters in calibration studies
(e.g.~\cref{fig:AllSrcSSMS}).  The overall
normalization is allowed to float within the estimated systematics errors
(8.6\%).  The background-PDF amplitudes \emph{within} the ROI are also
allowed to vary within their estimated systematic error (10.9\%).  The
\betascale{} is not allowed to float during the fit, but is manually profiled
while performing the PL scan for \nonubb{}. 

\begin{figure}
    \centering
	\includegraphics[width=0.98\columnwidth]{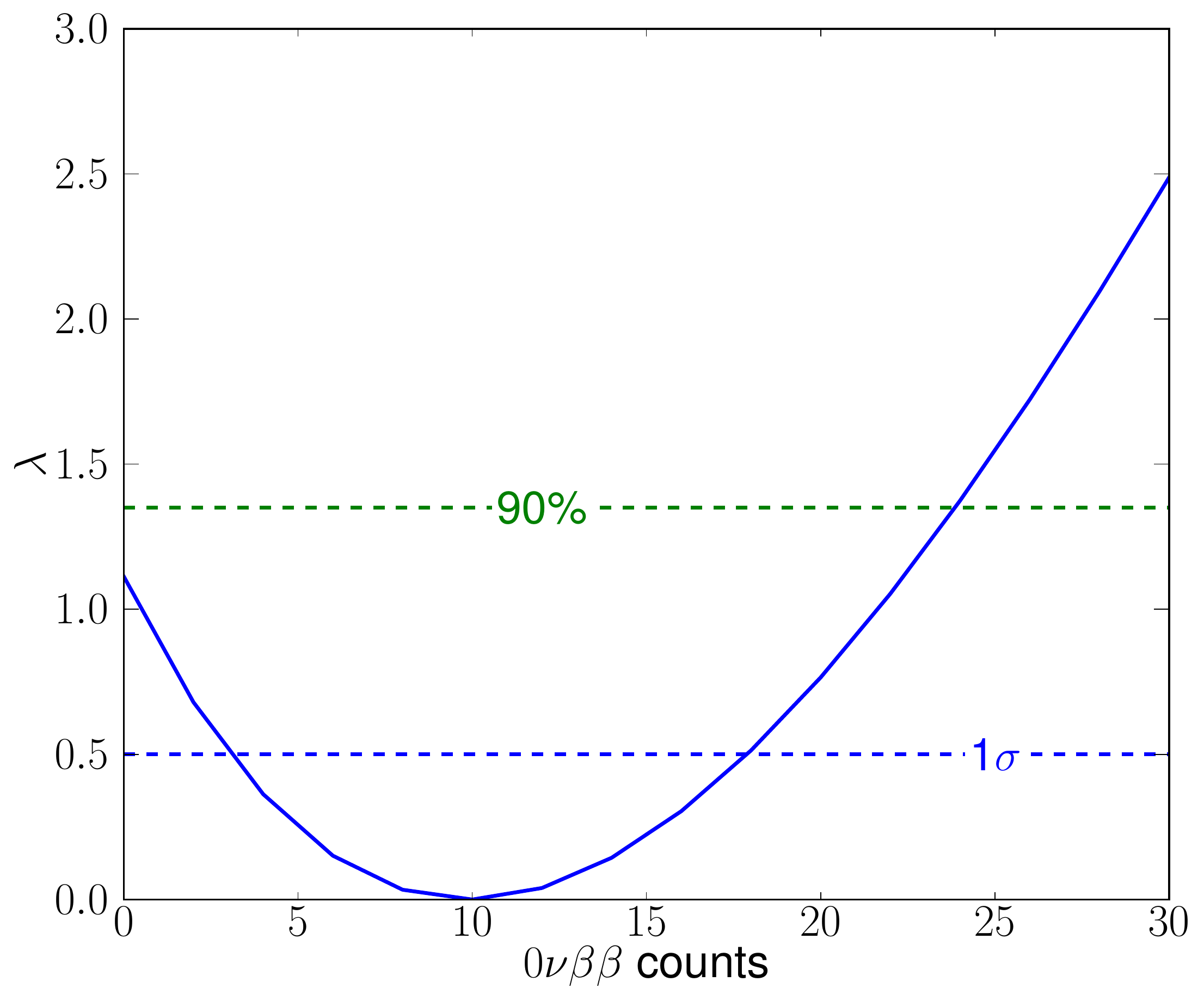} \caption{
\textbf{Profile likelihood, $\boldsymbol{\lambda}$, for \nonubbbf{} counts.}  The horizontal dashed lines
represent the $1\sigma$ and 90\% confidence levels assuming the validity of
Wilks' theorem~\cite{wilks1938,cowan1998statistical}, intersecting the profile
curve at $(3.1, 18)$ and 24 \nonubb{} counts, respectively.  From toy Monte
Carlo studies, the best-fit value is consistent with the null hypothesis at
$1.2\sigma$.  
}

\label{fig:Profile}
\end{figure}

The final step before performing the fit was the unmasking of live-time around
the SS ROI.  However, before unmasking the full data set, we investigated
backgrounds associated with Xe feeds, irregular occurrences in which additional
Xe gas is introduced into the purification circulation loop.  (These Xe feeds
occurred 10 times over the run period and are known to temporarily elevate, for
example, Rn levels in the detector.)  The live-time in the
two-week periods following the 10 feed events were unmasked first to search for
increased background levels in the ROI. No evidence for such an increase was
found and the unmasking of the remaining live-time proceeded.

The results of the ML fit are presented in \cref{fig:BestFit}.  The measured
\twonubb{} decay rate is consistent with~\cite{Albert2013}.  From the best-fit
model, the estimate of the background in the \nonubb{} $\pm2\sigma$ ROI is
$31.1\pm1.8 \mathrm{(stat)}\pm3.3 \mathrm{(sys)}$ counts, or $(1.7\pm0.2) \cdot
10^{-3}$~\kevkgyr{} normalized to the \emph{total} Xe exposure (123.7
kg$\cdot$yr).  Both this and the $\pm1\sigma$ value (also $(1.7\pm0.2) \cdot
10^{-3}$~\kevkgyr{}) are consistent with previous results, $1.5\pm0.1$
($1.4\pm0.1$) with the same units in the $\pm1\sigma$ ($\pm2\sigma$)
ROI~\cite{Auger:2012ar}.  
The
dominant backgrounds arise from \isot{Th}{232} (16.0 counts), \isot{U}{238}
(8.1
counts) and \isot{Xe}{137} (7.0 counts).  This amount of \isot{Xe}{137} is
consistent with estimates from studies of the activation of \isot{Xe}{136} in
muon-veto-tagged data.  The total number of events seen in this region is 39.
The best-fit value of \nonubb{} counts is 9.9, consistent with the null
hypothesis at $1.2\sigma$ as calculated using toy Monte Carlo studies.  The
corresponding PL scan of this parameter is shown in \cref{fig:Profile}.

A number of cross checks were performed on the result.  No event reconstruction
anomalies were found after hand-scanning all events in the ROI.  The
time-between-events distribution of the ROI events is consistent with a
constant-rate process and the SD distribution of events in data is consistent
with the best-fit model.  Additional backgrounds were considered that could
contribute events to the ROI.  In particular, we tested for \isot{Ag}{110m} and
\isot{Y}{88} because of their possible association with the measurement
in~\cite{Gando:2012zm}, and found that both produce a distinct high-multiplicity
signature in EXO-200 (SS/(SS+MS)\cussim{5-10\%}).  Separate fits
including each of these PDFs contributed the following counts to the $\pm2\sigma$
ROI: \numspec{Ag}{110m}{0.04\pm0.02} and \numspec{Y}{88}{0.02\pm0.01}.
Finally, we were able to exclude any significant effect on the ROI background
from \isot{Bi}{214} external to the Pb shield, e.g.~from \isot{U}{238} in the
surrounding salt. 

\begin{figure}
    \centering
	\includegraphics[width=0.98\columnwidth]{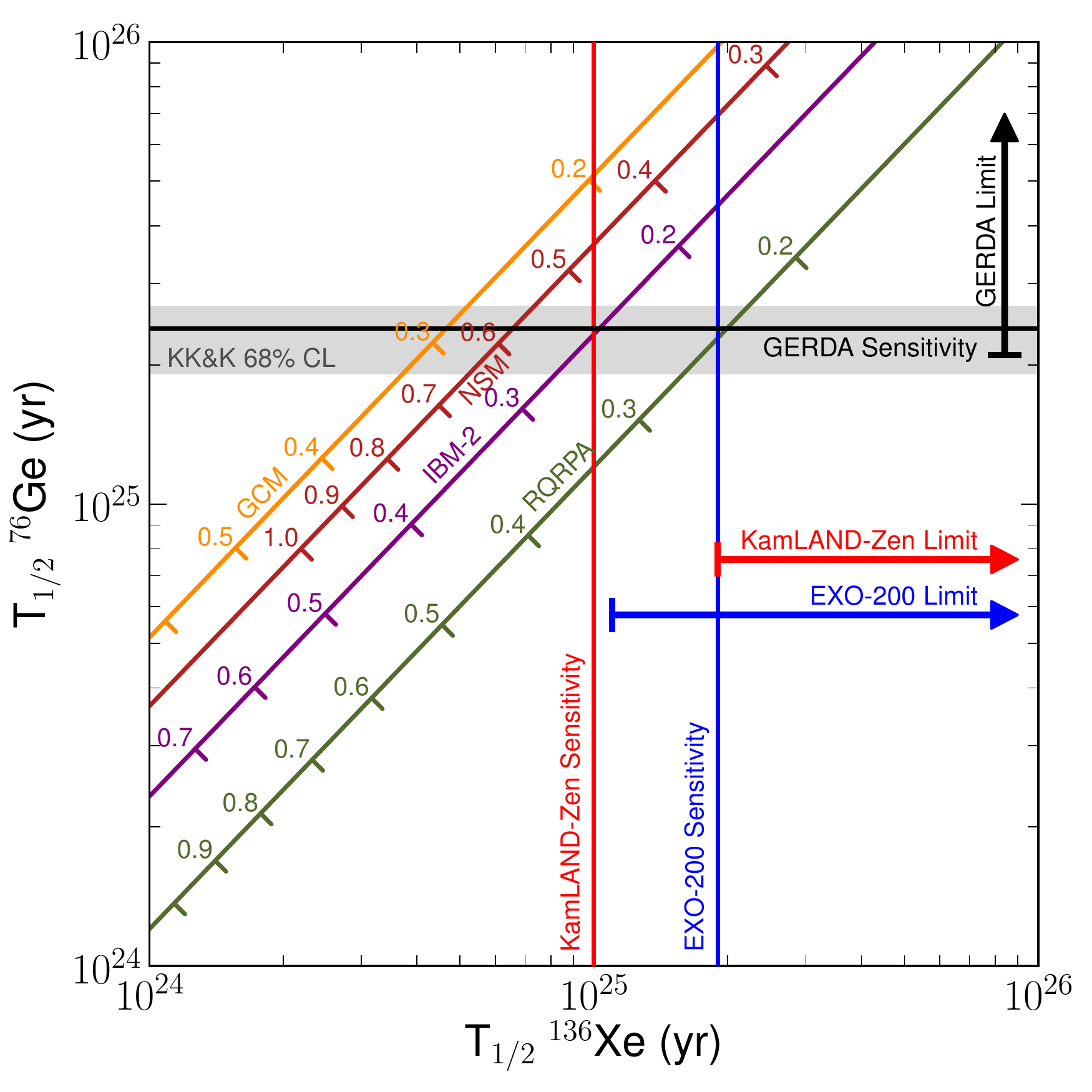} \caption{
\textbf{Comparison with recent results from \isotbold{Xe}{136} and \isotbold{Ge}{76}
\nonubbbf{} experiments.}  Sensitivity (orthogonal lines) and limits (arrows) from GERDA
and KamLAND-Zen are from~\cite{Agostini:2013mzu} and~\cite{Gando:2012zm},
respectively.  The diagonal lines are derived from several recent nuclear matrix
element calculations and the phase-space factor from~\cite{Kotila:2012zza},
included to allow comparison between results from the two nuclei:
GCM~\cite{Rodr:2010}, NSM~\cite{Menendez2009139}, IBM-2~\cite{Barea:2013}, and
RQRPA~\cite{Simkovic:2013}.  Tick marks along these lines indicate the
associated effective neutrino mass in eV.  The claimed observation in
\isot{Ge}{76} (KK\&K,~\cite{KlapdorKleingrothaus:2006ff}) is shown as a shaded
gray band.  The previous EXO-200 limit and sensitivity from~\cite{Auger:2012ar}  
were \halflife{1.6} and \halflife{0.7}, respectively.
}

\label{fig:SensCompare}
\end{figure}

\section{Discussion}

In summary, we report a 90\% C.L. lower limit on the \nonubb{} half-life of
\halflife{1.1}.  With the nuclear matrix elements
of~\cite{Rodr:2010,Menendez2009139,Barea:2013,Simkovic:2013} and phase space
factor from~\cite{Kotila:2012zza}, this corresponds to an upper limit on the
Majorana neutrino mass of 190\textendash{}450~meV.  Using the three flavor fit
of~\cite{Forero:2012} (with private communication, M.~Tortola, J.~Valle) we further use this range of effective
mass limits to construct a constraint on the mass $m_{\mathrm{min}}$ of the
lightest neutrino mass eigenstate, assuming the most disadvantageous
combination of CP phases. This corresponds to
$m_{\mathrm{min}}<0.69$\textendash{}1.63~eV, in case neutrinos are Majorana
particles. 

The results reported here supersede those of~\cite{Auger:2012ar}, owing to the
increased exposure and improved analysis.  The limit presented is however not
as strong as the limit from~\cite{Auger:2012ar}, consistent with expected
statistical fluctuations in the data.  An appropriate metric to
characterize the improvement of the experiment and independent of such
fluctuations is the `sensitivity', defined as the median expected 90\% CL
half-life limit assuming the background estimated from the ML fit and the
absence of a \nonubb{} signal.  We calculate this metric using an ensemble of
limits determined from Monte Carlo pseudo-experiments and find the EXO-200
sensitivity to be \halflife{1.9}, representing a factor of 2.7 improvement in
comparison to~\cite{Auger:2012ar}.

In \cref{fig:SensCompare} we compare the \nonubb{} sensitivity and half-life
limits from the GERDA, KamLAND-Zen, and EXO-200 experiments. Also shown is the
positive observation claim in \isot{Ge}{76}
from~\cite{KlapdorKleingrothaus:2006ff}. The results of the present analysis
are inconsistent with the central value of this claim at 90\% CL for two of the
four considered nuclear matrix element calculations: GCM~\cite{Rodr:2010} and
NSM~\cite{Menendez2009139}.

The first two years of EXO-200 data demonstrate the power of a large and
homogeneous LXe TPC in the search for \nonubb{}.  Simulations of the nEXO
experiment, a proposed 5000~kg LXe TPC based on the EXO-200 design,
show that the state-of-the-art background measured in EXO-200 can be further
improved by finer charge readout pitch (to improve the SS/MS discrimination)
and by lower electronic noise in the scintillation channel.
In addition Xe self-shielding will become more powerful in larger detectors,
where the $\gamma$ attenuation length at energies near the Q-value becomes
small with respect to the linear size of the LXe vessel.  This advantage
only applies to monolithic, homogeneous detectors.
\begin{acknowledgments}
	EXO-200 is supported by DOE and NSF in the United States, NSERC in Canada,
SNF in Switzerland, NRF in Korea, RFBR (12-02-12145) in Russia and DFG Cluster
of Excellence ``Universe'' in Germany. EXO-200 data analysis and simulation
uses resources of the National Energy Research Scientific Computing Center
(NERSC), which is supported by the Office of Science of the U.S. Department of
Energy under Contract No.~DE-AC02-05CH11231. The collaboration gratefully
acknowledges the WIPP for their hospitality.

\end{acknowledgments}

\section{Author Contributions}

Each of the authors of this article participated in the collection
and analysis of the data reported here, with the following exceptions: D.\ Beck
has contributed to the slow controls system; G.F.\ Cao has performed energy
resolution simulations; X.S.\ Jiang and Y.B.\ Zhao provide electronics expertise;
M.\ Danilov, A.\ Dolgolenko, T.\ Koffas, and P.\ Vogel contributed to the initial
conception and design of the experiment; M.\ Danilov and A.\ Dolgolenko also
contributed to the acquisition of the xenon, while P.\ Vogel also advises on
nuclear and particle theory; J.\ Davis, R.\ Nelson, and A.\ Rivas provide
engineering, operations, and technical support at the WIPP facility; A.\ 
Johnson, J.J.\ Russell, and A.\ Waite support data acquisition, data
processing, and software. Per collaboration policy, the authors are listed here
alphabetically. EXO-200 was constructed and commissioned by the authors of
Ref.~\cite{Auger:2012gs,Ackerman:2011gz}.

\bibliography{exo_Run2_0nubb}

\end{document}